# Quantum Time and Spatial Localization:
# An Analysis of the Hegerfeldt Paradox*

Francis S. G. Von Zuben[†]

Department of Physics and Astronomy, Texas Christian University, Fort Worth, Texas

12 May 2000

Two related problems in relativistic quantum mechanics, the apparent superluminal propagation of initially localized particles and dependence of spatial localization on the motion of the observer, are analyzed in the context of Dirac's theory of constraints. A parametrization invariant formulation is obtained by introducing time and energy operators for the relativistic particle and then treating the Klein-Gordon equation as a constraint. The standard, physical Hilbert space is recovered, via integration over proper time, from an augmented Hilbert space wherein time and energy are dynamical variables. It is shown that the Newton-Wigner position operator, being in this description a constant of motion, acts on states in the augmented space. States with strictly positive energy are non-local in time; consequently, position measurements receive contributions from states representing the particle's position at many times. Apparent superluminal propagation is explained by noting that, as the particle is potentially in the past (or future) of the assumed initial place and time of localization, it has time to propagate to distant regions without exceeding the speed of light. An inequality is proven showing the Hegerfeldt paradox to be completely accounted for by the hypotheses of subluminal propagation from a set of initial space-time points determined by the quantum time distribution arising from the positivity of the system's energy. Spatial localization can nevertheless occur through quantum interference between states representing the particle at different times. The non-locality of the same system to a moving observer is due to Lorentz rotation of spatial axes out of the interference minimum.

[†] Present address: Lockheed Martin Aeronautics Company, P.O. Box 748, Fort Worth, Texas 76101. Electronic mail address: vonzuben@ieee.org.

## I. INTRODUCTION

It is well known that there are problems in relativistic quantum mechanics regarding issues of spatial localization and causality. Although it has been generally acknowledged since the paper by Newton and Wigner[1] that positive energy states exist which are spatially localized at a particular time, Fleming,[2] Hegerfeldt,[3-6] and others[7] have shown that such states do not remain localized under time evolution; an effect which has been dubbed[8] the "Hegerfeldt Paradox." Its simplest demonstration is with the relativistic Klein-Gordon particle with a free positive energy Hamiltonian

$$H(\hat{\vec{p}}) = c\left(\hat{\vec{p}}^2 + m^2c^2\right)^{1/2} , \quad (1.1)$$

where $\hat{\vec{p}} = (\hat{p}^1, \hat{p}^2, \hat{p}^3)$ is the momentum, $c$ the speed of light, and $m$ the rest mass. If we take a strictly localized state $|\psi\rangle$ and translate it along $\vec{r}$ under the translation operator $\hat{U}(0, \vec{r}) = e^{-i\hat{\vec{p}}\cdot\vec{r}/\hbar}$, there exists some minimum $r$ such that $\hat{U}(0, \vec{r})|\psi\rangle$ is orthogonal to $|\psi\rangle$. But if $\hat{U}(0, \vec{r})|\psi\rangle$ evolves under the evolution operator $\hat{U}(ct, 0) = e^{-i\hat{H}t/\hbar}$ for any time $t$, the scalar product of $|\psi\rangle$ and $\hat{U}(ct, 0)\hat{U}(0, \vec{r})|\psi\rangle$ is

$$\langle\psi|\hat{U}^\dagger(ct, \vec{r})|\psi\rangle = \frac{c}{2\hbar}\int_{-\infty}^{+\infty}\frac{d^3p}{\omega(\vec{p})}\exp\left(\frac{i}{\hbar}\vec{p}\cdot\vec{r}\right)\exp\left[\frac{i}{\hbar}\left(\hat{\vec{p}}^2 + m^2c^2\right)^{1/2}ct\right]|\psi(\vec{p})|^2 , \quad (1.2)$$

where $\omega(\vec{p}) = H(\vec{p})/\hbar$ is the frequency. According to a theorem on the analyticity of Fourier transforms,[9,10] (1.2) cannot vanish because its Fourier transform is not an entire function; the exponential in $\left(\vec{p}^2 + m^2c^2\right)^{1/2}$ has singularities (branch points) on the complex hyperplane at the roots of $\vec{p}^2 = -m^2c^2$. Thus $\hat{U}(ct, \vec{r})|\psi\rangle$ is not orthogonal to $|\psi\rangle$ for any $\vec{r}$ and any $t \neq 0$. An apparent consequence is that the particle can be found outside the light cone of the initial locality, which is inconsistent with special relativity.

A similar result is suggested by the systems' phase velocities. Let $k^i = p^i/\hbar$ be the wave number along some axis; then the phase velocity is $\omega/k^i = c\sqrt{\vec{p}^2 + m^2c^2}/p^i$. Its magnitude is always greater than $c$, again implying superluminal dispersion of the wave packet. But the apparent conclusion that the particle actually travels faster than light contradicts well understood theorems[11] identifying the particle's velocity with the *group velocity* $d\omega/dk_i = cp^i/\sqrt{\vec{p}^2 + m^2c^2}$ , whose magnitude is always *less than c*.

Yet another problem is revealed by the theorem on Fourier transforms cited above. Even states which are localized at a particular instant of time are so in only one frame of reference.[1] They have infinite spatial extent in any reference frame moving relative to that frame. In the example just given, at $t = 0$ eq. (1.2) vanishes for some *r*. But if we transform $|\psi\rangle$ and $\hat{U}(ct, \vec{r})|\psi\rangle$ into a reference frame moving relatively with velocity $\vec{v}$ in the direction of $\vec{r}$ and calculate their scalar product at $t = 0$, we obtain

$$\langle\tilde{\psi}|\hat{U}^\dagger(0, \vec{r})|\tilde{\psi}\rangle = \frac{c}{2\hbar}\int_{-\infty}^{+\infty} \frac{d^3\tilde{p}}{\omega(\tilde{\vec{p}})}$$

$$\times \exp\left(\frac{i}{\hbar}\gamma\tilde{\vec{p}}\cdot\vec{r}\right)\exp\left[\frac{i}{\hbar}\gamma(\tilde{\vec{p}}^2 + m^2c^2)^{1/2}\vec{\beta}\cdot\vec{r}\right]\left|\psi\left[\gamma\left(\tilde{\vec{p}} + \frac{\vec{\beta}}{c}\hbar\omega(\tilde{\vec{p}})\right)\right]\right|^2 , \quad (1.3)$$

where the tilde ~ symbol indicates quantities transformed to the new frame of reference; $\vec{\beta} \equiv \vec{v}/c$ ; and $\gamma \equiv (1 - \vec{\beta}^2)^{-1/2}$. Again, (1.3) cannot vanish for any $\vec{r}$ because of the exponential in $(\tilde{\vec{p}}^2 + m^2c^2)^{1/2}$. Apparently spatial localization is not only an extraordinarily fleeting condition, but one which depends on the motion of the observer.

These problems have received various interpretations,[8,12-14] most commonly that the notion of spatial localization is not well defined in relativistic quantum mechanics: because a localizing potential, or the measurement interaction may cause particle/anti-particle pair creation, and the indistinguishability of resulting particles renders localization meaningless, it is argued that a satisfactory relativistic description must include indefinite number states, *i.e.*, quantum field theory.[8,13] But a free particle has no localizing potential which could induce the transitions between particle number states corresponding to particle creation or annihilation. The interaction of the position measurement itself might be thought do so, but no reference to this interraction is found in the problems cited above; it is not included in the theory. In any case, difficulties with the concept of localization are not restricted to cases involving a position operator. The Einstein-Podolsky-Rosen paradox[15,16] can be formulated without explicit position operators, as can Fermi's two-atom system,[17] which has recently been reevaluated by Hegerfeldt[5] and others.[18] For this and other reasons, Fleming[14] and others[19] stress the need to resolve these issues within quantum mechanics.

In this paper we propose a resolution to these problems based on the parametrization invariant formulation of quantum mechanics. The central idea is that the positive energy particle cannot be assigned a definite time, and problems arise from an unwarranted identification of *time of the measurement* with the *time of the particle*. The tacit assumption that these times must be the same is traceable to the strict interpretation in quantum mechanics of time as a parameter, not a dynamical variable.

In the standard theory a measurement with, say, the (Heisenberg picture) Newton-Wigner operator $\hat{\vec{Q}}(t)$ on a state $|\psi\rangle$ yielding result $\vec{r}$ is considered to imply the event of the particle being at the space-time point $(ct, \vec{r})$. Since per eq. (1.2) two consecutive measurements can yield two such results having space-like separation, superluminal propagation is inferred. But suppose the particle possessed, in addition to its three spatial variables $\hat{\vec{q}} = (\hat{q}^1, \hat{q}^2, \hat{q}^3)$, a quantum time variable $\hat{q}^0$. Then another interpretation is possible: $\vec{r}$ is the position of the particle extrapolated to *a particular value of the particle's time,* $q^0 = ct$, *ct* being the time of measurement[20,21] (in units of distance from the factor *c*). By "time of measurement" we mean the time on the experimenter's clock, assumed definite, when the measurment occurs. By "time of the particle" we refer to a quantum variable which is indefinite. Eigenvalues of $\hat{q}^0$ other than *ct* are allowed, corresponding to space-time points in the past and the future of $(ct, \vec{r})$. The locus of points arising from the particle's time uncertainty about the two measurement *times* includes pairs of space-time points associated with any two measurement *results* whose separation is *time-like*. Conclusions of superluminal propagation can thus be overcome.

This hypothesis raises another possibility: localization on a particular space-like hyperplane is an interference minimum arising from superposition of states having different values of $q^0$. *The non-locality of the same state viewed from another frame of* *reference is the result of Lorentz rotation of the spatial coordinate axes out of the hyperplane wherein the localized interference minimum occurs.*

Numerous quantum time formulations have been proposed (summaries may be found in refs. [22-24]). One which remains of interest, because of its manifest Lorentz covariance, is the proper time (or indefinite mass) formulation of quantum mechanics.[20,21,25] In the context of this theory, in which time in the rest frame is the parameter while time in other frames of reference is quantized, Horwitz and Usher show that states having equal distributions of positive and negative energy do not exhibit the Hegerfeldt paradox.[21] But a drawback of this description is the unphysical indefinite mass, implied by the canonical time operator.[19,26] However, definite masses may be recovered through the Dirac theory of constraints.[27,28] The resulting formulation has been called parametrization (or reparametrization) invariant,[29] because the action integral is unchanged regardless of which time-like variable serves as the parameter.

The parametrization invariant formulation will be employed in this paper. For simplicity, we will focus on the free Klein-Gordon particle; however, our results apply also to higher spin particles and particles in an external potential, as will be discussed in Sec. III. In Sec. II the parametrization invariant formulation is developed, yielding a description of a constrained system wherein observables appear as constants of the motion. A physical Hilbert space is derived, via integration over the proper time, from an augmented Hilbert space in which time and energy are dynamical variables. The Newton-Wigner operator is then derived naturally from the classical position observable. In Sec. III these results are applied to reveal the Hegerfeldt paradox and the non-Lorentz invariance of localized states to be consequences of the time indeterminacy of positive energy states. The main result of this paper is an inequality showing that the quantum time distribution of the particle entirely accounts for the probability of finding that same particle outside the light-cone of the assumed initial time and place of measurement, without assuming superluminal velocities. Conclusions are given in Sec. IV.

## II.  THE PARAMETRIZATION INVARIANT FORMULATION

In order to make this paper self-contained and introduce notation, we will briefly review Dirac constraint theory as applied to the relativistic particle.  Dirac showed that, for any physical system, it is possible to take the time parameter as an *additional coordinate*, introduce a *new parameter* to track system progress, and then impose the physical constraints which are implied by the formulation.[27]  When quantized, the physical states are just those of the standard non-constrained theory; however, the description suggests a freedom in time which the standard theory does not recognize.

In constraint theory, constraints are classified as first or second class according to whether or not they Poisson-commute with all other constraints.  Systems having only first class constraints can be quantized by well established procedures.  This method was originally developed Dirac[27] and Bergmann[28] during the 1950's, having as one objective a quantum theory of gravity.[23,29]  It has since found numerous applications, especially in field theories.  See refs. [30,31] for further discussions.

### A.  Classical Description of the Free Particle as a Constrained System

We take as a starting point[32] a Lagrangian $L = -mc\left(c^2 - \vec{v}^2\right)^{1/2}$ dependent on three spatial velocities $v^i \equiv dq^i/dt$, from which the usual Hamiltonian is obtained by computing the conjugate momenta $p_i \equiv \partial L/\partial v^i$ and performing the Legendre transformation $H \equiv p_i v^i - L$ (repeated indices imply summation, Latin indices running over the three spatial axes).  The resulting expression, when quantized, is eq. (1.1).

#### i. The Dirac Hamiltonian

To obtain a parametrization invariant description, we will take *ct* as an *additional coordinate* $q^0$, introduce a *new parameter* $\tau$ (which is real-valued and has units of time), and define a new Lagrangian $L_S$ in terms of *L*,

$$L_S \equiv \left(\frac{dt}{d\tau}\right)L = -mc\left(\dot{q}_\mu \dot{q}^\mu\right)^{1/2} . \tag{2.1}$$

$L_S$ is manifestly Lorentz covariant, depending on four canonical velocities $\dot{q}^\mu \equiv dq^\mu/d\tau$ (throughout this paper the over-dot means differentation with respect to $\tau$, not $t$; repeated Greek indices imply summation over the four space-time axes; and the metric tensor is diagonal with elements $g_{\mu\mu} = +1, -1, -1, -1$ ). The canonical momenta are defined

$$p_\mu \equiv -\frac{\partial L_S}{\partial \dot{q}^\mu} = \frac{mc\dot{q}_\mu}{\left(\dot{q}_\nu \dot{q}^\nu\right)^{1/2}} . \tag{2.2}$$

The minus sign in the definition of the momenta is determined by choice of metric and relativistic sign conventions.[33,34] Eq. (2.2) yields, besides the three spatial momenta $p_i$ already defined in the previous paragraph, an additional temporal momentum $p_0$.

Since the dependence of $t$ on $\tau$ is not yet specified, the denominator $\left(\dot{q}_\nu \dot{q}^\nu\right)^{1/2} = (dt/d\tau)\left(c^2 - \vec{v}^2\right)^{1/2}$ of (2.2) contains the arbitrary factor $dt/d\tau$. Therefore, the four momenta $p_\mu$ are not uniquely defined in terms of the velocities $\dot{q}_\mu$, and the system is "singular" in the sense of ref. [35] (hence the subscript $S$). Furthermore, the four momenta are not all independent of each other; from eq. (2.2) the sum of the squared momenta is $m^2c^2$. We recognize this as a *primary, first class constraint,* since it is implied by $L_S$, and (as will be shown) it Poisson-commutes with other constraint(s) to be identified below. We formally write this constraint as follows:

$$\varphi \equiv p_\mu p^\mu - m^2c^2 \approx 0 . \tag{2.3}$$

Constraint (2.3) will become the Klein-Gordon equation when quantized. The symbol $\approx$ denotes a weak equality, which, in constraint theory terminology means $\varphi$ is not set to zero until *after* any Poisson brackets in a given expression have been calculated.[27] Physically, this constraint reflects the fact that not all values of the four momenta are accessible to the system; $p$ is constrained to hyperboloid (2.3). We may rewrite eq. (2.3) as $p_0 \approx \pm H/c$, *i.e.*, the energy is proportional to the standard Hamiltonian.

The Lagrangian $L_S$ is homogenous of the first degree in the velocities $\dot{q}^u$. Consequently, from Euler's theorem on homogeneous functions,[35] a vanishing canonical

Hamiltonian is obtained, defined by the Legendre transformation

$$H_C \equiv -p_\mu \dot{q}^\mu - L_S = 0 \tag{2.4}$$

[the minus sign preceding $p_\mu \dot{q}^\mu$ follows from that in eq. (2.2)]. The Dirac Hamiltonian, denoted $H_D$, is defined as the sum of the canonical Hamiltonian $H_C$, and each primary first class constraint multiplied by an undetermined multiplier $\lambda$. Since $H_C$ vanishes, and there is (as yet) only one constraint, $H_D$ consists *entirely* of the constraint $\varphi$ multiplied by $\lambda$. We may replace $\lambda$ with $\lambda/2m$ to provide a form of $H_D$ analogous to a non-relativistic Hamiltonian with constant potential, thus obtaining

$$H_D = \frac{\lambda}{2m} \varphi = \frac{\lambda}{2m} \left( p_\mu p^\mu - m^2 c^2 \right) . \tag{2.5}$$

In analyzing the Hegerfeldt paradox we shall be concerned with the propagation of the particle from one space-time point $x'(\tau') = \left(x'^0, \vec{x}'\right)$ to another $x''(\tau'') = \left(x''^0, \vec{x}''\right)$. The total action for this propagation is

$$J(\tau', \tau'') = \int_{t'}^{t''} dt\, L = \int_{\tau'}^{\tau''} d\tau \frac{dt}{d\tau} L = \int_{\tau'}^{\tau''} d\tau\, L_S \approx \int_{\tau'}^{\tau''} d\tau \left( -p_\mu \dot{q}^\mu - H_D \right) . \tag{2.6}$$

From (2.6) we note that, as long as $x'^\mu(\tau') = x'^\mu(t')$ and $x''^\mu(\tau'') = x''^\mu(t'')$, the action $J$ is invariant under the reparametrization $t \to \tau$.[23] It is also proportional to the length of the particle's world-line, since $dJ = -mc \left( g_{\mu\nu} dq^\mu dq_\nu \right)^{1/2}$.[36] Between endpoints, the trajectory follows from the requirement that $J$ be stationary, leading, from the last equality in eq. (2.6), to Hamilton's equations of motion. These equations will be written with the aid of the fundamental Poisson bracket relations,

$$\left[ q^\mu, p_\nu \right]_P = \delta^\mu_{\ \nu} , \tag{2.7a}$$

$$\left[ q^\mu, q^\nu \right]_P = \left[ p_\mu, p_\nu \right]_P = 0 , \tag{2.7b}$$

where $\delta^\mu_{\ \nu} = 1, \mu = \nu; 0, \mu \neq \nu$ is the Kronicker delta, and $\left[ A, B \right]_P \equiv \left( \partial A/\partial q^\mu \right)\left( \partial B/\partial p_\mu \right) - \left( \partial B/\partial q^\mu \right)\left( \partial A/\partial p_\mu \right)$ is the Poisson bracket. Hamilton's equations of motion are then:

$$\dot{q}^\mu = \left[ q^\mu, H_D \right]_P = \frac{\lambda}{2m} \left[ q^\mu, \varphi \right]_P + \frac{1}{2m} \left[ q^\mu, \lambda \right]_P \varphi \approx \lambda \frac{p^\mu}{m} , \tag{2.8a}$$

$$\dot{p}_\mu = \left[ p_\mu, H_D \right]_P = \frac{\lambda}{2m} \left[ p_\mu, \varphi \right]_P + \frac{1}{2m} \left[ p_\mu, \lambda \right]_P \varphi \approx 0 . \tag{2.8b}$$

Comparison of (2.8a) and (2.2) reveals that $\lambda = \left( \dot{q}_\nu \dot{q}^\nu \right)^{1/2}/c$; in other words, the multiplier λ is a function of the velocities, carried into the Hamiltonian formalism because the momenta are not uniquely determined.[27] Since λ may vary arbitrarily and still yield a Hamiltonian such that the system remains on the constraint hypersurface (2.3), it is a gauge variable without physical significance. In fact, a well-known feature of the parametrization invariant formulation is that "motion" generated by the Dirac Hamiltonian $H_D$ is indistinguishable from gauge transformations.[23,29,30,31,34]

### ii. Gauge Fixation and the Definition of Observables

In order to fix the gauge, we place a new constraint directly on λ,[23,34] specifically requiring that λ be a constant of the motion through the condition

$$\varphi_g \equiv \dot{\lambda} = \frac{\partial \lambda}{\partial \tau} + \left[\lambda, H_D\right]_P = \frac{\partial \lambda}{\partial \tau} + \frac{\lambda}{2m} \frac{\partial \lambda}{\partial q^\mu} p^\mu \approx 0 \ . \qquad (2.9)$$

From (2.9) we may write, using (2.8a) and (2.2),

$$dq_\mu dq^\mu - \left[c\, d(\lambda \tau)\right]^2 \approx 0 \ . \qquad (2.10)$$

*Therefore, we identify λτ with the proper time.* This gauge is accordingly referred to as the *proper time gauge.* Since $\left[\varphi_g, \varphi\right]_P \approx 0$, both constraints φ and $\varphi_g$ are first class.

Now let us return to eqs. (2.8). Since (2.8b) vanishes weakly, the momenta $p_\mu$ are constants of the motion. On the other hand, (2.8a) does not vanish, and as a result λ appears in the final expression for $\dot{q}^\mu$. This makes $q^\mu(\tau)$ dependent on the gauge choice (2.9),[24,29,34] so the coordinates remain arbitrary. In gauge theories, observables must be *gauge invariant*,[30,31] and since in the parametrization invariant formulation $H_D$ is the generator of gauge transformations, *observables must poisson-commute weakly with $H_D$.* That is, *observables must be constants of the motion.* This requirement is not met by the coordinates $q^\mu$. But if we subtract from the *spatial* coordinate $q^i$ the *i*'th component of the three-velocity multiplied by the time coordinate $q^0$, we obtain a constant of the motion. Using (2.8a) and (2.2) we write

$$v^i = c\frac{dq^i}{dq^0} = c\frac{\dot{q}^i}{\dot{q}^0} = c\frac{p^i}{p^0} \ . \tag{2.11}$$

Thus, the quantity

$$Q^i = q^i - \frac{p^i}{p_0}q^0 \tag{2.12}$$

has weakly vanishing Poisson brackets with $H_D$ (as may be checked). Also, the three quantities $Q^i$ have weakly vanishing Poisson brackets with each other. The three-vector $\vec{Q} = \left(Q^1, Q^2, Q^3\right)$ will therefore be taken as an observable corresponding to the spatial position of the particle extrapolated to the time $q^0 = 0$.[37] In fact it is $Q^i$, not $q^i$, which when quantized is an Hermitian operator on the physical Hilbert space, where it is equivalent to the Newton-Wigner position operator.

### B. Quantum Description of the Free Particle as a Constrained System

Quantization is accomplished as in standard quantum mechanics by formally replacing the dynamical variables with operators, and replacing the Poisson bracket with the commutator divided by $i\hbar$. This leads to a Schrödinger equation involving the Dirac Hamiltonian and the parameter τ. As in the standard theory, expectation values correspond to classical quantities. To recover the physical description, constraints (2.3) and (2.9) are imposed on the system. But constraints cannot simply be written as relations between operators; for instance, we cannot write (2.3) as $\hat{p}_0 = \pm\sqrt{\hat{\vec{p}}^2 + m^2c^2}$, because this would eliminate $\hat{p}_0$ and $\hat{q}^0$ as independent variables, which is inconsistent with Poisson brackets (2.7). Instead, we recognize that operators have states corresponding to all mathematical values of the dynamical variables. We will refer to these generalized states as *augmented* states, and we denote the Hilbert space to which they belong as $\mathcal{H}_{aug}$. Constraints, on the other hand, are *conditions set on states*, consistent with their classical description as weak equalities [see eq. (2.3)]. States satisfying the constraint relations are referred to as *physical* states, belonging to the physcial Hilbert space $\mathcal{H}_{phy}$. Since the primary constraint is actually the Klein-Gordon equation, and physical states are solutions to this equation, the physcial states are the same as those of standard relativistic quantum mechanics.

### i. Quantization

We now proceed with quantization of the system. The Dirac Hamiltonian is

$$\hat{H}_D = \frac{\hat{\lambda}}{2m}\hat{\varphi} = \frac{\hat{\lambda}}{2m}\left(\hat{p}_\mu \hat{p}^\mu - m^2 c^2\right) , \qquad (2.13)$$

where the constraint $\hat{\varphi}$ and the multiplier $\hat{\lambda}$ are now shown as operators (quantum operators are designated with the caret ^). The fundamental commutation relations are

$$\left[\hat{q}^\mu, \hat{p}_\nu\right] = i\hbar \delta^\mu_{\ \nu} , \qquad (2.14a)$$

$$\left[\hat{q}^\mu, \hat{q}^\nu\right] = \left[\hat{p}_\mu, \hat{p}_\nu\right] = 0 , \qquad (2.14b)$$

where $\left[\hat{A}, \hat{B}\right] \equiv \hat{A}\hat{B} - \hat{B}\hat{A}$. The equations of motion (2.8) have the quantum form

$$\frac{d}{d\tau}\left\langle \psi(\tau) \left| \hat{A} \right| \psi(\tau) \right\rangle = \left\langle \psi(\tau) \left| \frac{\partial \hat{A}}{\partial \tau} + \frac{1}{i\hbar}\left[\hat{A}, \hat{H}_D\right] \right| \psi(\tau) \right\rangle , \qquad (2.15)$$

where in (2.15) $\hat{A}$ is an arbitrary function of $\hat{q}^\mu$, $\hat{p}_\mu$, or $\hat{\lambda}$. Equation (2.15) leads to the Schrödinger equation in the same way as in standard non-relativistic quantum mechanics,

$$i\hbar \frac{d}{d\tau}\left| \psi(\tau) \right\rangle = \hat{H}_D \left| \psi(\tau) \right\rangle , \qquad (2.16)$$

and has solutions of the form

$$\left| \psi(\tau) \right\rangle = \exp\left(\frac{-i}{\hbar}\hat{H}_D \tau\right) \left| \psi(0) \right\rangle . \qquad (2.17)$$

The proper-time constraint (2.9) is imposed as a condition on the states:

$$\left\langle \psi(\tau) \left| \hat{\varphi}_g \right| \psi(\tau) \right\rangle = \left\langle \psi(\tau) \left| \frac{\partial \hat{\lambda}}{\partial \tau} + \frac{1}{i\hbar}\left[\hat{\lambda}, \hat{H}_D\right] \right| \psi(\tau) \right\rangle = 0 . \qquad (2.18)$$

Finally, to meet the condition of gauge invariance as stated in Sec. II-A-ii, every observable $\hat{O}$ must commute with the constraint operator $\hat{\varphi}$:

$$\left[\hat{O}, \hat{\varphi}\right] = 0 . \qquad (2.19)$$

### ii. The Augmented Hilbert Space

The linear vector space of normalizable solutions of the Schrödinger equation

(2.16) comprises the *augmented Hilbert space* $\mathcal{H}_{aug}$. Operators $\hat{p}_\mu$, $\hat{q}^\mu$, and $\hat{\lambda}$ are assumed Hermitian on $\mathcal{H}_{aug}$,[38] and the scalar product on $\mathcal{H}_{aug}$ is

$$\langle \phi(\tau') | \psi(\tau) \rangle = \int_{-\infty}^{+\infty} d^4p \, \phi^*(p,\tau') \psi(p,\tau) = \int_{-\infty}^{+\infty} d^4q \, \phi^*(q,\tau') \psi(q,\tau) , \quad (2.20)$$

where coordinate-representation functions $\psi(q^0, q^1, q^2, q^3, \tau)$ are defined in terms of the Fourier transform, $\psi(q,\tau) = (2\pi\hbar)^{-4/2} \int_{-\infty}^{+\infty} d^4p \exp\left(\frac{-i}{\hbar} p_\mu q^\mu\right) \psi(p,\tau)$, of functions $\psi(p_0, p_1, p_2, p_3, \tau)$. The identity operator in $\mathcal{H}_{aug}$ is expressible in terms of orthonormal and complete sets of eigenvectors $|p_\mu\rangle$ and $|q^\mu\rangle$ of $\hat{p}_\mu$ and $\hat{q}^\mu$,

$$\langle p' | p \rangle = \delta^{(4)}(p - p') , \quad \langle q' | q \rangle = \delta^{(4)}(q - q') , \quad (2.21\text{a,b})$$

$$\hat{I} = \int_{-\infty}^{+\infty} d^4p \, |p\rangle\langle p| = \int_{-\infty}^{+\infty} d^4q \, |q\rangle\langle q| , \quad (2.22)$$

where $\delta(x)$ is the Dirac delta function (except where otherwise noted, basis vectors for $\mathcal{H}_{aug}$ are shown as $|g\rangle = |g^0\rangle \otimes |g^1\rangle \otimes |g^2\rangle \otimes |g^3\rangle$ ).

### iii. The Physical Hilbert Space

Unlike the augmented Hilbert space $\mathcal{H}_{aug}$, whose simple structure resembles non-relativistic quantum mechanics, the physical Hilbert space $\mathcal{H}_{phy}$, is complicated by constraints. To understand the Hegerfeldt paradox, it will be necessary in Sec. III to make use of expressions defined on *both* $\mathcal{H}_{aug}$ and $\mathcal{H}_{phy}$; that is, we will represent states and operators defined on $\mathcal{H}_{phy}$ in terms of cooresponding quantities defined on $\mathcal{H}_{aug}$. We shall therefore briefly discuss the mathematical relationship between the two Hilbert spaces. Further discussion may be found in refs. [24,34,39].

The constraint $\hat{\phi}$ determines the physical state vectors, which are solutions of

$$\langle q | \hat{\phi} | \psi \rangle = \langle q | \hat{p}_\mu \hat{p}^\mu - m^2c^2 | \psi \rangle = \left[ -\hbar^2 \frac{\partial^2}{\partial q^{02}} + \hbar^2 \nabla^2 - m^2c^2 \right] \psi(q) = 0 . \quad (2.23)$$

We recognize in (2.23) the Klein-Gordon equation, from which two observations are made: 1) physical states are eigenstates of $\hat{H}_D$ with eigenvalue zero (since $\hat{H}_D$ contains $\hat{\varphi}$ as a factor); and 2) physical states are solutions of the Klein-Gordon equation. From eq. (2.16), such states are independent of the parameter $\tau$; accordingly, physical states $|\psi\rangle$ will be distinguished notationally from augmented states $|\psi(\tau)\rangle$ simply by the absence of the parameter $\tau$ in their argument. Physical states are not generally normalizable in $\mathcal{H}_{aug}$,[29,31,39] and therefore belong to the separate Hilbert space $\mathcal{H}_{phy}$, the state space of standard relativistic quantum mechanics.

To derive the scalar product in $\mathcal{H}_{phy}$, we take the scalar product in $\mathcal{H}_{aug}$ of states $|x'(\tau')\rangle$ and $|x''(\tau'')\rangle$, then integrate over their proper-time difference $\hat{\lambda}(\tau'' - \tau')$:

$$\langle x''^0, \vec{x}'' \mid x'^0, \vec{x}'\rangle = (2m)^{-1} \int_{-\infty}^{+\infty} d(\tau'' - \tau')\langle x''(\tau'') | \hat{\lambda} | x'(\tau')\rangle$$

$$= \int_{-\infty}^{+\infty} d(\tau'' - \tau')\langle x''(0) | \hat{\lambda} \exp\left[i\,\hat{\lambda}\,\hat{\varphi}\,(\tau'' - \tau')\right] | x'(0)\rangle$$

$$= 2\pi\hbar \left\langle x''(0) \left| \delta\left[\hat{p}_\mu \hat{p}^\mu - m^2 c^2\right] \right| x'(0) \right\rangle . \qquad (2.24)$$

Note that the gauge constraint (2.18) allows us to eliminate $\hat{\lambda}$ in (2.24) (because $\hat{\lambda}$ commutes weakly with $\hat{\varphi}$). Equation (2.24) is a *sum over histories* for all paths connecting the space-time points $x' = (x'^0, \vec{x}')$ and $x'' = (x''^0, \vec{x}'')$. Since the duration of each history in terms of its proper time $\hat{\lambda}\tau$ is equal to the length of that history's path [see eq. (2.10)], and all histories contribute to the total transition amplitude, setting the integration range in (2.24) to $(-\infty, +\infty)$ yields the *total propagator* from $x'$ to $x''$ and *vice versa*.[23,24] The classical trajectory emerges through destructive interference between all paths except those for which the phase of the integrand is stationary.[40]

We may write eq. (2.24) in the momentum representation by inserting eq. (2.2):

$$\langle x''^0, \vec{x}'' \mid x'^0, \vec{x}'\rangle = (2\pi\hbar)^{-3} \int_{-\infty}^{+\infty} \frac{d^4 p}{2|p_0|} \exp\left[\frac{i}{\hbar} p_\mu (x''^\mu - x'^\mu)\right]$$

$$\times \left[\delta\left(p_0 - \sqrt{\vec{p}^2 + m^2 c^2}\right) + \delta\left(p_0 + \sqrt{\vec{p}^2 + m^2 c^2}\right)\right]$$

$$= (2\pi\hbar)^{-3} \int_{-\infty}^{+\infty} \frac{d^3p}{\sqrt{\vec{p}^2 + m^2c^2}} \exp\left[\frac{i}{\hbar}\vec{p}\cdot(\vec{x}'-\vec{x}'')\right] \cos\left[\frac{1}{\hbar c}H(\vec{p})(x'^0 - x''^0)\right], \quad (2.25)$$

where $H(\vec{p})$ is the standard Hamiltonian (1.1).[41] Equation (2.25) is the Hadamard (or Schwinger) Green's function,[42] which will be taken as the scalar product of *physical* states $|x'^0, \vec{x}'\rangle$ and $|x''^0, \vec{x}''\rangle$. These states may be written as $|x^0, \vec{x}\rangle \equiv |x^0, \vec{x}_+\rangle + |x^0, \vec{x}_-\rangle$, the sum of positive and negative energy states, which in turn may be written $|x^0, \vec{x}_\pm\rangle = e^{\mp\frac{i}{\hbar c}H(\vec{p})x^0}|\vec{x}\rangle$, where $|\vec{x}\rangle$ satisfies $\hat{\vec{q}}|\vec{x}\rangle = \vec{x}|\vec{x}\rangle$ (basis vectors for $\mathcal{H}_{phy}$ are shown as $|\vec{g}\rangle = |g^1\rangle \otimes |g^2\rangle \otimes |g^3\rangle$ unless noted otherwise). Note that the Lorentz invariant momentum-space measure $d^3p/\sqrt{\vec{p}^2+m^2c^2}$ [obtained in evaluating (2.25)] prevents $|x^0, \vec{x}\rangle$ and $|x^0, \vec{x}_\pm\rangle$ from being orthogonal for different values of $x$, consistent with the fact that $\hat{\vec{q}} = i\hbar\nabla_{\vec{p}}$ is not Hermitian on $\mathcal{H}_{phy}$.[43]

Equation (2.24) or (2.25) could serve as the starting point for a path integral formulation of relativistic quantum mechanics, and Halliwell and Ortiz observe[24] that the Hadamard Green's function (2.25) supports causal propagation. Furthermore, the author shows elsewhere[34] that the Hegerfeldt paradox can be explained by noting that the endpoints of path integrals representing the apparent non-causal propagation are separated by time-like intervals. In this paper, however, we shall adhere to the canonical formulation of relativistic quantum mechanics, but we will retain the negative-energy sector. The momentum eigenstates $|\vec{p}_\pm\rangle$ satisfying $\hat{\vec{p}}|\vec{p}_\pm\rangle = \vec{p}|\vec{p}_\pm\rangle$ and $\hat{p}_0|\vec{p}_\pm\rangle = \pm\sqrt{\vec{p}^2+m^2c^2}|\vec{p}_\pm\rangle$ form together an orthonormal complete basis for $\mathcal{H}_{phy}$:

$$\langle \vec{p}'_+|\vec{p}_+\rangle = \langle \vec{p}'_-|\vec{p}_-\rangle = 2\sqrt{\vec{p}^2+m^2c^2}\,\delta^{(3)}(\vec{p}-\vec{p}'); \quad \langle \vec{p}'_-|\vec{p}_+\rangle = 0; \quad (2.26\text{a,b})$$

$$\hat{I} = \int_{-\infty}^{+\infty} \frac{d^3p}{2\sqrt{\vec{p}^2+m^2c^2}}\left(|\vec{p}_+\rangle\langle\vec{p}_+| + |\vec{p}_-\rangle\langle\vec{p}_-|\right). \quad (2.27)$$

The scalar product in the momentum representation is therefore

$$\langle \phi|\psi\rangle = \int_{-\infty}^{+\infty} \frac{d^3p}{2\sqrt{\vec{p}^2+m^2c^2}}\left[\phi_+^*(\vec{p})\psi_+(\vec{p}) + \phi_-^*(\vec{p})\psi_-(\vec{p})\right], \quad (2.28)$$

where $\psi_\pm(\vec{p})$ are the positive and negative frequency portions of the function

$\psi(\vec{p}) = \psi_+(\vec{p}) + \psi_-(\vec{p})$. Defining the coordinate representation functions $\psi_\pm(q)$ as Fourier transforms $\psi_\pm(q) = (2\pi\hbar)^{-3/2} \int_{-\infty}^{+\infty} d^3p\, e^{\frac{i}{\hbar}[\vec{p}\cdot\vec{q} \mp H(\vec{p})q^0/c]} \psi_\pm(q) / 2\sqrt{\vec{p}^2 + m^2c^2}$ of $\psi_\pm(\vec{p})$, where $\psi(q) = \psi_+(q) + \psi_-(q)$ is the sum of the positive and negative frequency components, the scalar product in the coordinate representation in $\mathcal{K}_{phy}$ becomes

$$\langle \phi | \psi \rangle = \frac{i\hbar}{2} \int_{-\infty}^{+\infty} d^3q \left[ \phi_+^*(q) \overleftrightarrow{\frac{\partial}{\partial q^0}} \psi_+(q) - \phi_-^*(q) \overleftrightarrow{\frac{\partial}{\partial q^0}} \psi_-(q) \right], \qquad (2.29)$$

where the symbol $\overleftrightarrow{\partial/\partial x}$ is defined by $\phi(\overleftrightarrow{\partial/\partial x})\psi \equiv \phi(\partial\psi/\partial x) - (\partial\phi/\partial x)\psi$. Using coordinate eigenstates $|x^0, \vec{x}_\pm\rangle$ defined below eq. (2.25) (Fourier transforms of $|\vec{p}_\pm\rangle$), the identity operator for $\mathcal{K}_{phy}$ in the coordinate representation is

$$\hat{I} = \frac{i\hbar}{2} \int_{-\infty}^{+\infty} d^3q \left( |q^0, \vec{q}_+\rangle \overleftrightarrow{\frac{\partial}{\partial q^0}} \langle q^0, \vec{q}_+| - |q^0, \vec{q}_-\rangle \overleftrightarrow{\frac{\partial}{\partial q^0}} \langle q^0, \vec{q}_-| \right). \qquad (2.30)$$

The unusual resolution of (2.30), with negative frequency terms subtracted, accounts for the scalar product being positive definite despite inclusion of negative energies.[24]

For a given $\hat{\lambda}$, the time coordinate $\hat{q}^0$ may move forward or backward according as $\hat{p}_0$ is positive or negative [see eq. (2.8a)]. Negative energies correspond to the antiparticle. Since the product $p_0 \dot{q}^0$ does not change sign, the action, eq. (2.6), is unchanged by transitions $\pm p_0$. Note that $c^{-1}(\hat{p}_\mu \hat{p}^\mu)^{1/2}$ is a mass operator, implying that quantum states in this formulation have indefinite mass; however, discrete mass eigenvalues $\pm m$ are recovered on physical states with $+m$ yielding the decreasing action integral (2.6).

### iv. Representation of Physical States in Terms of Augmented States

We conclude this subsection by showing how an arbitrary physical state may be represented in terms of augmented states. The key to this representation is integration of the augmented states over the proper time. For an arbitrary physical state $|\psi\rangle$, there exist augmented states $|\psi(\tau)\rangle$ such that[34,39]

$$|\psi\rangle = (2\pi m\hbar)^{-1}\int_{-\infty}^{+\infty} d(\hat{\lambda}\tau)\hat{N}|\psi(\tau)\rangle = \pi^{-1}\int_{-\infty}^{+\infty} d\tau\, e^{-i\hat{\varphi}\tau}\hat{N}|\psi(0)\rangle = 2\delta(\hat{\varphi})\hat{N}|\psi(0)\rangle \ . \quad (2.31)$$

$\hat{N}$ is a normalizing factor chosen to permit $|\psi\rangle$ and $|\psi(\tau)\rangle$ to be normalized respectively in $\mathcal{H}_{phy}$ and $\mathcal{H}_{aug}$ (*i.e.*, $\langle\psi(\tau)|\psi(\tau)\rangle = \langle\psi|\psi\rangle = 1$). We require that $[\hat{N},\hat{\varphi}] = 0$. The resulting state $|\psi\rangle$ satisfies constraint (2.23), and is thus physical. An example of this representation for Newton-Wigner states is given below in eq. (2.36).

We now make an important observation about observables acting on physical states. Since an observable $\hat{O}$ commutes with constraint $\hat{\varphi}$, per eq. (2.19), we may write

$$\hat{O}|\psi\rangle = \pi^{-1}\int_{-\infty}^{+\infty} d\tau\exp(-i\hat{\varphi}\tau)\hat{O}\left[\hat{N}|\psi(0)\rangle\right] . \quad (2.32)$$

In other words, we may *first* move $\hat{O}$ under the integral to the right of $e^{-i\hat{\varphi}\tau}$, *operate on the corresponding augmented state*, and then integrate over the proper time.

### C. Spatially Localized States and the Position Observable

It was noted below eq. (2.25) that the physical states $|x^0,\vec{x}\rangle$ and $|x^0,\vec{x}_\pm\rangle$ are not orthogonal for different values of *x* because of the factor $(\vec{p}^2 + m^2c^2)^{-1/2}$ appearing in the physical momentum space measure. Localized states such as Newton-Wigner states can be orthogonal because they are proportional to $(\vec{p}^2 + m^2c^2)^{1/4}$; this renders the scalar product an entire function, but spoils the states' Lorentz invariance, localization occurring in only one frame of reference. In the remainder of this paper it is understood, unless stated otherwise, that the discussion pertains to that particular frame of reference.

#### i. Newton-Wigner States

To incorporate Newton-Wigner states into our formalism, we wish to represent them terms of augmented states, as in eq. (2.31). Since $\mathcal{H}_{phy}$ includes states with positive and negative energy, while Newton-Wigner states are restricted to either positive or negative energy, there is room for two orthogonal copies of each physical position state for every $\vec{x}$. To restrict the states which we shall construct to a single energy sector, we will first define a basis of positive (negative) energy *augmented* states as follows:

$$|x_\pm(\tau)\rangle \equiv \theta(\pm \hat{p}_0)|x(\tau)\rangle , \qquad (2.33)$$

where $|x(\tau)\rangle$ is an eigenstate of $\hat{q}$ at $\tau = 0$ with eigenvalue $x$, and we insert the unit step function $\theta$ to eliminate negative (positive) energies.

Now consider a property of states $|x_+(\tau)\rangle$: *they are non-local in time*, since their scalar product with $|q(\tau)\rangle$ is

$$\langle q(\tau) | x_+(\tau)\rangle = (2\pi\hbar)^{-4} \int_{-\infty}^{+\infty} d^3p \, \exp\left[\frac{i}{\hbar}\vec{p}\cdot(\vec{x}-\vec{q})\right] \int_{-\infty}^{+\infty} dp_0 \, \theta(p_0) \exp\left[\frac{-i}{\hbar}p_0(x^0-q^0)\right]$$

$$= \delta^{(3)}(\vec{q}-\vec{x}) f(q^0-x^0) , \qquad (2.34)$$

where

$$f(q^0-x^0) = \lim_{\varepsilon \to +0} (2\pi)^{-1} \frac{i}{q^0-x^0+i\varepsilon} = \frac{i}{2\pi}P\left(\frac{1}{q^0-x^0}\right) + \frac{1}{2}\delta(q^0-x^0) , \qquad (2.35)$$

P denoting Caucy principle part. The function $f$, the Fourier transform of the unit step function $\theta(p_0)$, is not the delta function, implying that $|x_+(0)\rangle$ does not represent a single point in space-time. The *time coordinates represented by* $|x_+(0)\rangle$ *are distributed over* $(-\infty, +\infty)$, although their amplitudes are strongly peaked at $q^0 = x^0$, as illustrated in Fig. 1. This time indeterminacy, an example of the time-energy uncertainty relation, will be seen to be responsible for the Hegerfeldt paradox.

We now define the positive (negative) energy Newton-Wigner states $|x^0, \vec{x}_{nw\pm}\rangle$:

$$|x^0, \vec{x}_{nw\pm}\rangle \equiv (2\pi m\hbar)^{-1} \int_{-\infty}^{+\infty} d(\hat{\lambda}\tau) \left(2\pi\hbar\hat{p}_0\right)^{1/2} |x_\pm(\tau)\rangle . \qquad (2.36)$$

Comparing with eq. (2.31), the factor $\hat{N}$ is $\left(2\pi\hbar\hat{p}_0\right)^{1/2}$. Taking $x^0 = 0$ in (2.36) we obtain, upon integration over $\tau$, $\langle p | 0, \vec{x}_{nw\pm}\rangle = \left(\vec{p}^2 + m^2c^2\right)^{1/4} e^{-i\vec{p}\cdot\vec{x}/\hbar}$, the conventional Newton-Wigner state.[1] We also define the physical states

$$|x^0, \vec{x}_{loc}\rangle \equiv 2^{-1/2}|x^0, \vec{x}_{nw+}\rangle + 2^{-1/2}|x^0, \vec{x}_{nw-}\rangle , \qquad (2.37)$$

equal superpositions of positive and negative energy Newton-Wigner states, which are truly local (in the Hegerfeldt sense); hence the subscript *loc*.

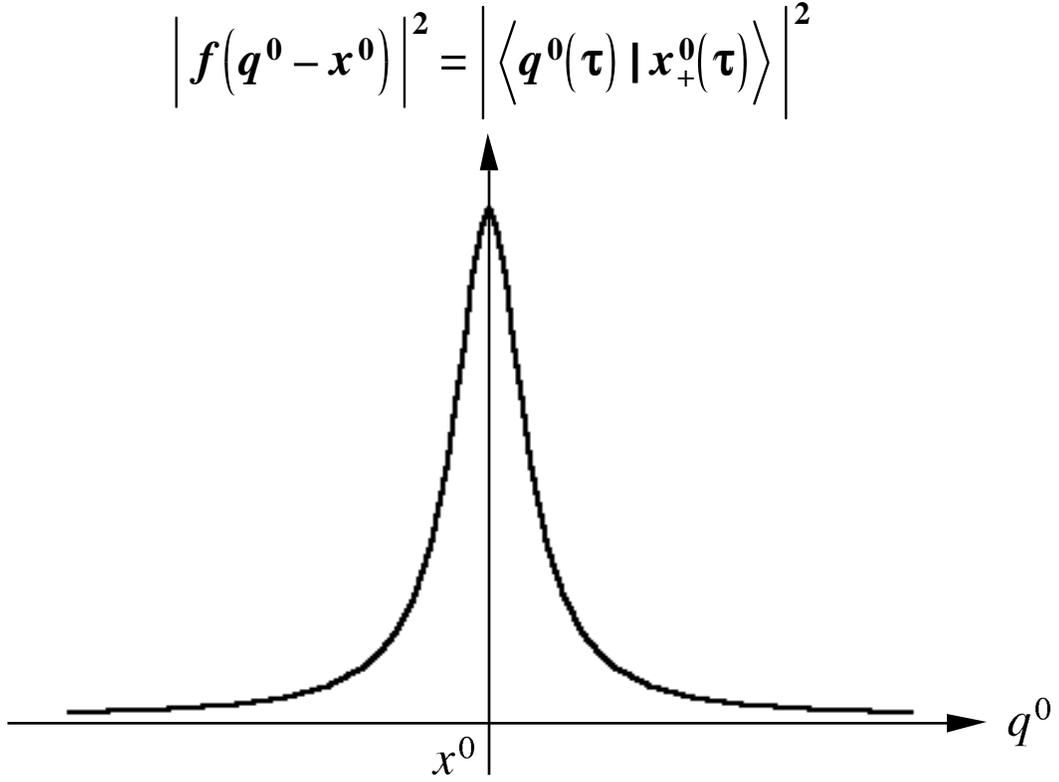

Figure 1. Quantum time distribution of the positive energy particle. This graph represents the augmented state $\left|x_+(\tau)\right\rangle \equiv \theta(\hat{p}_0)\left|x(\tau)\right\rangle$ at $\tau = 0$, the wave function for which is given by eq. (2.35) (the limit $\varepsilon \to +0$ is not yet taken). Restriction to only positive frequency results, through the Fourier transformation, in an indefinite time coordinate, which leads to the localization paradoxes of corresponding physical states constructed from augmented states $\left|x_+(\tau)\right\rangle$. The line shape is that of the Lorentzian from quantum optics.

Scalar products for $|x^0, \vec{x}_{loc}\rangle$ and $|x^0, \vec{x}_{nw\pm}\rangle$ in $\mathcal{H}_{phy}$ are:

$$\langle x'^0, \vec{x}'_{nw\pm} | x^0, \vec{x}_{nw\pm}\rangle = (2\pi\hbar)^{-3} \int_{-\infty}^{+\infty} d^3p \, \exp\left[\mp\frac{i}{\hbar c}H(\vec{p})(x^0 - x'^0)\right] \exp\left[\frac{i}{\hbar}\vec{p}\cdot(\vec{x}-\vec{x}')\right],$$

$$\langle x'^0, \vec{x}'_{nw+} | x^0, \vec{x}_{nw-}\rangle = 0, \qquad (2.38\text{a,b,c})$$

$$\langle x'^0, \vec{x}'_{loc} | x^0, \vec{x}_{loc}\rangle = (2\pi\hbar)^{-3} \int_{-\infty}^{+\infty} d^3p \, \cos\left[\frac{1}{\hbar c}H(\vec{p})(x^0 - x'^0)\right] \exp\left[\frac{i}{\hbar}\vec{p}\cdot(\vec{x}-\vec{x}')\right].$$

Both eq. (2.38a) and (2.38c) become the delta function for $x'^0 = x^0$. It will be noted that, unlike (2.38a), eq. (2.38c) is the Fourier transform of $\cos\left[\sqrt{\vec{p}^2 + m^2c^2}\,(x^0 - x'^0)\right]$, which is analytic on the entire complex plane. This may be seen from the expansion $\cos(x) = 1 - x^2/2! + x^4/4! + \ldots$ involving only even powers of the exponent, eliminating the branch point at the roots of $\vec{p}^2 = -m^2c^2$. As noted by Horwitz and Usher,[21] states may therefore be constructed in terms of states $|x^0, \vec{x}_{loc}\rangle$ which do not exhibit the instantaneous spreading cited by Hegerfeldt. Finally, we note that states $|x^0, \vec{x}_{loc}\rangle$ form a complete set on $\mathcal{H}_{phy}$, i.e.,

$$\hat{I} = \int_{-\infty}^{+\infty} d^3q \, |q^0, \vec{q}_{loc}\rangle\langle q^0, \vec{q}_{loc}|. \qquad (2.39)$$

### ii. The Newton-Wigner Operator

As noted in Sec. II-A, observables for the Klein-Gordon particle in the parametrization invariant formulation are constants of the motion. In particular, the classical position observable $Q^i$ defined in eq. (2.12) is a constant of motion, and has the meaning of the position of the particle extrapolated to the time $q^0 = 0$. To find the quantum operator for $Q^i$, we follow the arguments that lead us to eq. (2.12); however, for more generality we extrapolate to the time $\hat{q}^0 = ct$. From now on we shall treat $ct$ as a *particular* value of the operator $\hat{q}^0$ ($ct$ is therefore a number, not an operator).

As the time coordinate changes by $\Delta q^0$, the spatial position changes by $(d\vec{q}/dq^0)\Delta q^0 = \vec{v}\Delta q^0$, where $\vec{v}$ is the spatial velocity. O'Connel and Wigner have shown[11] that, in spite of the Hegerfeldt paradox with its apparent implications regarding superluminal velocity, the spatial velocity operator $\hat{v}^i$ is nevertheless $c\hat{p}^i/\hat{p}^0$. Thus, if we subtract $(\hat{q}^0 - ct)(\hat{p}^i/\hat{p}_0)$ from $\hat{q}^i$, we will have "backed out" all motion since the time $\hat{q}^0 = ct$, *i.e.*, we will obtain a constant of the motion. But since $\hat{p}_0$ does not commute with $\hat{q}^0$, it is necessary to use a symmetrized product involving the anti-commutator, $\{\hat{q}^0, \hat{p}_0^{-1}\}$, where $\{\hat{A}, \hat{B}\} \equiv \hat{A}\hat{B} + \hat{B}\hat{A}$. This yields

$$\hat{Q}^j(t) \equiv \hat{q}^j - \frac{1}{2}\hat{p}^j\{(\hat{q}^0 - ct), \hat{p}_0^{-1}\} . \tag{2.40}$$

Evaluating the anti-commutator, (2.40) becomes in the momentum representation

$$\hat{Q}^j(t) = i\hbar\frac{\partial}{\partial p_j} - \frac{p^j}{p_0}\left(i\hbar\frac{\partial}{\partial p_0} - ct\right) + \frac{i\hbar p^j}{2(p_0)^2} . \tag{2.41}$$

The corresponding three-vector $\hat{\vec{Q}}(t) = \left[\hat{Q}^1(t), \hat{Q}^2(t), \hat{Q}^3(t)\right]$ is the position of the particle extrapolated to $\hat{q}^0 = ct$, as was originally found by Horwitz and Piron in the context of the theory of indefinite mass.[20]

The observable $\hat{\vec{Q}}(t)$ obeys the following commutation relations:

$$\left[\hat{Q}^j, \hat{p}_k\right] = i\hbar\delta^j{}_k , \tag{2.42a}$$

$$\left[\hat{Q}^j, \hat{Q}^k\right] = 0 , \tag{2.42b}$$

$$\left[\hat{Q}^j, \hat{H}_D\right]|\psi\rangle = \frac{i\hbar\hat{\lambda}}{m}\left(-\hat{p}_j + \frac{\hat{p}_j}{\hat{p}_0}\hat{p}_0\right)|\psi\rangle = 0 , \tag{2.42c}$$

where the commutator with $\hat{H}_D$ is shown multiplied into a physical state, allowing us to drop the commutator with $\hat{\lambda}$ [*i.e.*, $\left[\hat{Q}^j, \hat{H}_D\right]$ vanishes weakly; compare eqs. (2.8)].

The states $|x^0, \vec{x}_{nw\pm}\rangle$ and $|x^0, \vec{x}_{loc}\rangle$ are eigenfunctions of $\hat{\vec{Q}}(t = x^0/c)$ with eigenvalue $\vec{x}$ [see eq. (3.18) below]. It can be shown[20,21,34] that $\hat{\vec{Q}}(t)$ is Hermitian on $\mathcal{K}_{phy}$. Also, on the constraint hypersurface $\varphi = 0$, $\hat{\vec{Q}}(t)$ is just the Newton-Wigner position operator in the Heisenberg picture, *i.e.*,

$$\hat{Q}^j(t)\bigg|_{\varphi=0} = e^{\pm\frac{i}{\hbar}H(\vec{p})t}\left[i\hbar\frac{\partial}{\partial p_j} + \frac{i\hbar p^j}{2(\vec{p}^2 + m^2c^2)}\right]e^{\mp\frac{i}{\hbar}H(\vec{p})t} . \tag{2.43}$$

We conclude this section by deriving two expansions for $\hat{\vec{Q}}(t)$ as integral expressions of dyads $|a\rangle O_{ab}\langle b|$, $O_{ab}$ being matrix elements, which will be employed in

Sec. III. If such an expansion is in a representation for which $\hat{O}$ is diagonal and states $|a\rangle$ are orthonormal, it is a spectral decomposition, yielding the probability amplitude $\langle a | \psi \rangle$ of obtaining the eigenvalue $O_{aa}$ for an arbitrary state $|\psi\rangle$. Evidently from (2.38) and (2.39), such a representation is provided by the states $|x^0, \vec{x}_{loc}\rangle$ which form an orthonormal basis for $\mathcal{H}_{phy}$. A spectral decomposition is therefore

$$\hat{Q}^j(t)\Big|_{\varphi=0} = \int_{-\infty}^{+\infty} d^3q \, |ct, \vec{q}_{loc}\rangle \vec{q} \, \langle ct, \vec{q}_{loc}| \, . \tag{2.44}$$

Expressions like $\int d^n x \, |x\rangle x \langle x|$ in eqs. (2.44) and (2.45) (below) are to be read as $\int dx^1 \, |x^1\rangle x^1 \langle x^1| \otimes \int dx^2 \, |x^2\rangle x^2 \langle x^2| \otimes \ldots \otimes \int dx^n \, |x^n\rangle x^n \langle x^n|$. Since (2.45) is constructed in terms of physical states, it is a *constrained operator,* generally valid only if not commuted with other operators. If we wish to obtain an expansion consistent with the Dirac constraint methodology, $\hat{\tilde{Q}}(t)$ must be represented by *augmented states.* Using commutators (2.14), the expression $\hat{\tilde{v}} = c\hat{\vec{p}}/\hat{p}^0$ for velocity, and an expansion of the form $\int_{-\infty}^{+\infty} d^4q \, |q\rangle q \langle q|$ in place of $\hat{q}$, we obtain

$$\hat{\tilde{Q}}(t) = \int_{-\infty}^{+\infty} d^4q \, |q(\tau)\rangle (1,\vec{q}) \langle q(\tau)| - \hat{\tilde{v}} \int_{-\infty}^{+\infty} d^4q (\hat{p}_0)^{1/2} |q(\tau)\rangle \left(\frac{q^0}{c} - t, \vec{1}\right) \langle q(\tau)| (\hat{p}_0)^{-1/2} , \tag{2.45}$$

where the states $|q^\mu(\tau)\rangle$ are eigenstates of $\hat{q}^\mu$ at $\tau = 0$. Eq. (2.45) is not a spectral decomposition (since it is not diagonal), but it exhibits the spectral properties of the position observable in terms of the particle's intrinsic quantum time $\hat{q}^0$.

### III. QUANTUM TIME AND SPATIAL LOCALIZATION

In this section we propose a resolution to the spatial localization problems discussed in Sec. I. For simplicity, we will begin with a positive energy Klein-Gordon particle state prepared to definitely yield the value $\vec{q} = (0, 0, 0) = \vec{0}$ when measured at time $q^0 = 0$ with the Newton-Wigner operator $\hat{\tilde{Q}}(0)$; in other words, the Newton-Wigner state $|0, \vec{0}_{nw+}\rangle$. The Hegerfeldt paradox suggests this system can be found at position $\vec{q} = \vec{r}$ at time $q^0 = ct$, with $|\vec{r}| > |ct|$. We will show the reason for this is simply that $|0, \vec{0}_{nw+}\rangle$ represents the particle located at $\vec{0}$ at different times $q^0$, including times such that $(ct, \vec{r})$ is within the light-cone of $(q^0, \vec{0})$.

First we will evaluate the expression $\hat{\vec{Q}}(t)\left|0, \vec{0}_{nw+}\right\rangle$ which represents this measurement in the Dirac constraint formalism of Sec. II. Then we will present our main result: an inequality showing that the quantum time distribution of $\left|0, \vec{0}_{nw+}\right\rangle$ completely accounts for the positive probability distribution of the particle outside the light-cone of $(0, \vec{0})$, with the particle traveling at subluminal velocities. We will finally show that the reason positive energy Newton-Wigner states are localized at all, in view of their time distribution, is destructive interference between contributions from different times. This interference minimum does not occur in other frames of reference, which is why localization is not Lorentz invariant. We conclude the section with remarks generalizing these results to other particle states, and a discussion of implications for causality.

## A. Analysis of a Position Measurement

Let us evaluate the expression $\hat{\vec{Q}}(t)\left|0, \vec{0}_{nw+}\right\rangle$, which represents a position measurment, at a measurement time $t$, of the Newton-Wigner state localized at the origin at time zero. We will express $\left|0, \vec{0}_{nw+}\right\rangle$ as an expansion of augmented states as in eqs. (2.31) and (2.36), then move the operator $\hat{\vec{Q}}(t)$ under the integral sign as in eq. (2.32):

$$\hat{\vec{Q}}(t)\left|0, \vec{0}_{nw+}\right\rangle = \hat{\vec{Q}}(t)\left[(2\pi m\hbar)^{-1}\int_{-\infty}^{+\infty} d(\hat{\lambda}\tau)\,(2\pi\hbar\hat{p}_0)^{1/2}\left|0_+(\tau)\right\rangle\right]$$

$$= (2\pi)^{-1/2}\int_{-\infty}^{+\infty} d\tau \exp(-i\hat{\phi}\tau)\,\hat{\vec{Q}}(t)\,(\hat{p}_0)^{1/2}\left|0_+(0)\right\rangle \qquad (3.1)$$

Now we substitute for $\hat{\vec{Q}}(t)$ the expansion (2.45) which represents the operator in terms of augmented states $|q(\tau)\rangle$:

$$\hat{\vec{Q}}(t)\left|0, \vec{0}_{nw+}\right\rangle = (2\pi)^{-1/2}\int_{-\infty}^{+\infty} d\tau \exp(-i\,\hat{\phi}\,\tau) \qquad (3.2)$$

$$\times\left[\int_{-\infty}^{+\infty} d^4q\,|q(0)\rangle(1,\vec{q})\langle q(0)| - \hat{\vec{v}}\int_{-\infty}^{+\infty} d^4q\,(\hat{p}_0)^{1/2}|q(0)\rangle\left(\frac{q^0}{c}-t, \vec{1}\right)\langle q(0)|\right]\left\|0_+(0)\right\rangle .$$

Finally we multiply the expressions inside the square brackets into the augmented state $|0_+(0)\rangle$ [see eq. (2.34)], and perform the integration over $\tau$ to obtain

$$\hat{\vec{Q}}(t)|0, \vec{0}_{nw+}\rangle = \int_{-\infty}^{+\infty} dq^0\, \hat{\vec{v}}\left(t - \frac{q^0}{c}\right) f(q^0)|q^0, \vec{0}_{loc}\rangle \, , \tag{3.3}$$

where $f$ is defined in (2.35).

Eq. (3.3) is a superposition of physical states $|q^0, \vec{0}_{loc}\rangle$ localized at the origin at different times $q^0$, each multiplied by an amplitude $f(q^0) = i\mathrm{P}(1/2\pi q^0) + \delta(q^0)/2$, each multiplied in turn by a position eigenvalue $\vec{v}(t - q^0/c)$. In other words, eq. (3.3) predicts a spread of possible measurement results $\vec{v}(t - q^0/c)$ whose uncertainty arises from two sources: 1) the velocity $\vec{v} = c\vec{p}/p_0$ of the particle is indeterminate yielding different position values when multiplied by a time difference $t - q^0/c$ ; 2) *the time $q^0$ of the particle is indeterminate, yielding a range of differences $t - q^0/c$ between the time of the particle and the time of measurement.*

This situation is illustrated in Fig. 2, where we exhibit the time axis and one spatial axis passing through the space-time origin. Shading indicates regions for which $\langle q^0, q^1{}_{loc}|0, 0_{nw+}\rangle$ does not vanish. Besides the origin $(0, 0)$ and measurement event $(ct, r)$, a point in the past $(x'^0, 0)$ and another in the future $(x''^0, 0)$ are also plotted. Forward and backwards light cones are shown as dashed lines. The vector $(ct, r)$ is shown as space-like, making this an illustration of the Hegerfeldt paradox.

It is apparent from eq. (3.3) that the measurement receives contributions from the past, including some time $x'^0$ such that $(x'^0, 0)$ has $(ct, r)$ within its light-cone. But, in order to propagate from $(x'^0, 0)$ to $(ct, r)$ (dashed arrow) *the particle does not exceed the speed of light.* It does have to pass through the $q^0 = 0$ line where it has zero probability of being found (save at $q^1 = 0$ ). We return to this point in Sec. III-C.

Notice that the states $|q^0, \vec{0}_{loc}\rangle$ in the expansion on the right side of eq. (3.3) are not positive energy Newton-Wigner states, but the *positive-negative energy superpositions defined in eq. (2.37).* The negative energy components represent (in relativistic quantum mechanics) backward-in-time motion.  Thus, the measurement also receives contributions from the *future*.  This contribution includes some time $x'''^0$ such that $(ct, r)$ is within the backward light-cone of $(x'''^0, 0)$.  Again, to propagate from $(x'''^0, 0)$ to $(ct, r)$ (dashed arrow) the particle does not exceed the speed of light.

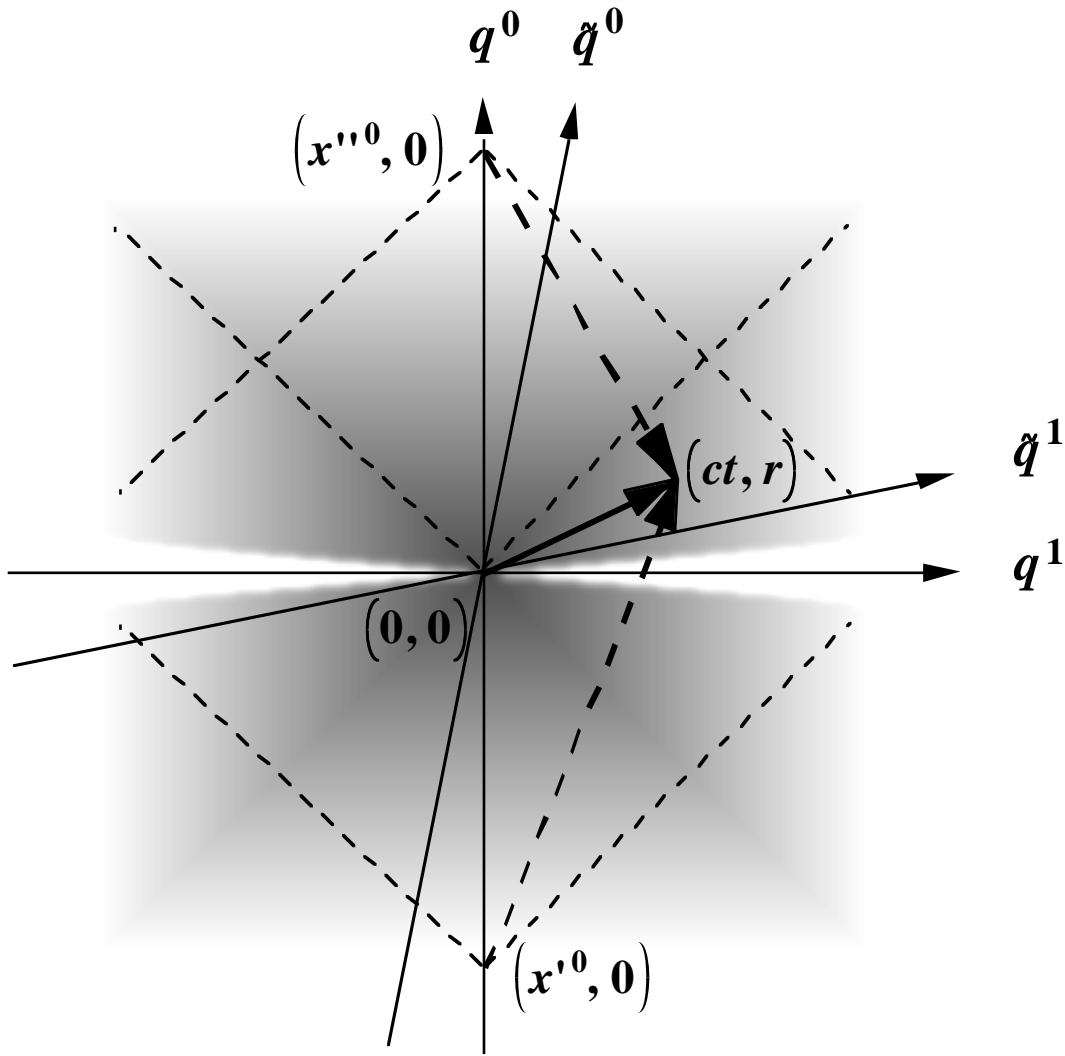

Figure 2. Minkowski diagram of Hegerfeldt scenario. Shading notionally depicts regions where the Newton-Wigner function $\langle q^0, q^1{}_{loc} | 0, 0_{nw+}\rangle$ is non-vanishing. Both proper and moving (denoted with a tilde ~) coordinate axes are shown. Light dashed lines denote forward or backward light cones. Apparent superluminal propagation from $(0, 0)$ to $(ct, r)$ is the result of subluminal propagation from earlier points such as $(x'^0, 0)$ or later points such as $(x''^0, 0)$. Localization on the $q^0 = 0$ plane is an interference minimum which the moving observer does not see because his spatial axis is tilted out of the plane where the minimum occurs.

If the time of the particle represented by $|0, \vec{0}_{nw+}\rangle$ were definitely $q^0 = 0$, as is generally tacitly assumed, the measurement $\hat{\vec{Q}}(t)$ could yield values only within the light cone of $(0, \vec{0})$.  This is because the speed is bounded by $|c\vec{p}/p_0| \approx |c\vec{p}/\sqrt{\vec{p}^2 + m^2c^2}| < c$ (since the states acted on by the velocity operator $\hat{\vec{v}}$ are physical), confining possible measurement result events $(ct, \vec{v}t)$ to time-like distances from $(0, \vec{0})$.  In fact, this is precisely the result obtained when we substitute $|0, \vec{0}_{loc}\rangle$ in place of $|0, \vec{0}_{nw+}\rangle$ in eq. (3.1), as has already been shown by Horwitz and Usher.[21] However, since the positive energy Newton-Wigner state $|0, \vec{0}_{nw+}\rangle$ represents the particle a times other than $q^0 = 0$, $\hat{\vec{Q}}(t)$ can and will yield values outside the light-cone.

We wish to point out that the statement, "the measurement receives contributions from the past or future," simply means that, if the particle is at the origin at time $q^0/c$ and is moving with constant velocity $\vec{v}$, then when the particle's time is $t$, its position will be $\vec{v}(t - q^0/c)$.  Or, an equivalent way of saying "the particle is at position $\vec{r}$ at time $t$" is to say "the particle is on a world line which passes through the point $(ct, \vec{r})$."

### B. The Space-Time Inequality

The question arises, if the time uncertainty of the particle represented by $|0, \vec{0}_{nw+}\rangle$, surrounding the time $q^0 = 0$, explains its propagation to a space-time point $(ct, \vec{r})$ outside the light cone of $(0, \vec{0})$ at subluminal velocities, should it not be possible to establish that the probability of the particle's position being $\vec{q} = \vec{0}$, summed over all times $q^0$ such that $(ct, \vec{r})$ becomes accessible within the light cone of $(q^0, \vec{0})$, is greater than the probability of the particle being found at all points which are a distance $r$ from the origin at time $t$; and could we not thereby entirely account for the distribution outside the light cone of $(0, \vec{0})$ ?

In fact, the answer to the above question is yes. Refer to Fig. 3. For those times $q^0 < r - ct$, $(ct, r)$ is within the light-cone of $(q^0, 0)$ (bottom dotted arrow). For times $q^0 > r + ct$, $(ct, r)$ is within the backward light cone of $(q^0, 0)$ (top dotted arrow). Since probabilities are additive, it is sufficient to show that the probability density $\mathscr{P}(q^0 = r - ct)$ for $q^0$ being equal to $r - ct$, plus the probability density $\mathscr{P}(q^0 = r + ct)$ for $q^0$ being equal to $r + ct$, is always greater than the probability density $\mathscr{P}(t, |\vec{q}| = r)$ that a measurement at time $t$ will find the particle at a distance $r$ from the origin, given that $t > 0$ and $r > ct$. If this is established, then upon integration of $\mathscr{P}(q^0 = r - ct) + \mathscr{P}(q^0 = r + ct)$ over all $r$ in the interval $(ct, +\infty)$, the result is greater than the total probability of a measurement finding the particle outside the light cone $r = \pm ct$ at time $t$, obtained by integrating $\mathscr{P}(t, |\vec{q}| = r)$ over the same interval $(ct, +\infty)$. This would suffice to entirely account for the particle's positive probability distribution outside the light cone of $(0, \vec{0})$ with subluminal velocities. Therefore, we need to prove that

$$\mathscr{P}(q^0 = r - ct) + \mathscr{P}(q^0 = r + ct) \geq \mathscr{P}(t, |\vec{q}| = r) . \qquad (3.4)$$

Eq. (3.4) will be referred to as the *space-time inequality*.

From eqs. (2.34) and (2.35), the particle represented by $|0, \vec{0}_{nw+}\rangle$ has a *time distribution*, described by the augmented wave function

$$f(q^0) = \langle q^0(\tau) | 0_+(\tau) \rangle = \lim_{\varepsilon \to +0} \frac{1}{2\pi} \frac{i}{(q^0 + i\varepsilon)} . \qquad (3.5)$$

The distribution $f$ is normalized (in the sense of distributions) because

$$\int_{-\infty}^{+\infty} dq^0 f(q^0) = \lim_{\gamma \to +0} \frac{i}{2\pi} \oint_C dq^0 \frac{\exp(-i\gamma q^0)}{q^0 + i\varepsilon} = 1 , \qquad (3.6)$$

where the contour $C$ is taken clockwise around the pole at $q^0 = -i\varepsilon$, enclosing the lower half-plane, yielding a residual of unity. A further remark about normalizing (3.5) is given below eq. (3.12). We will treat $f(q^0)$ as the probability amplitude for the time of the

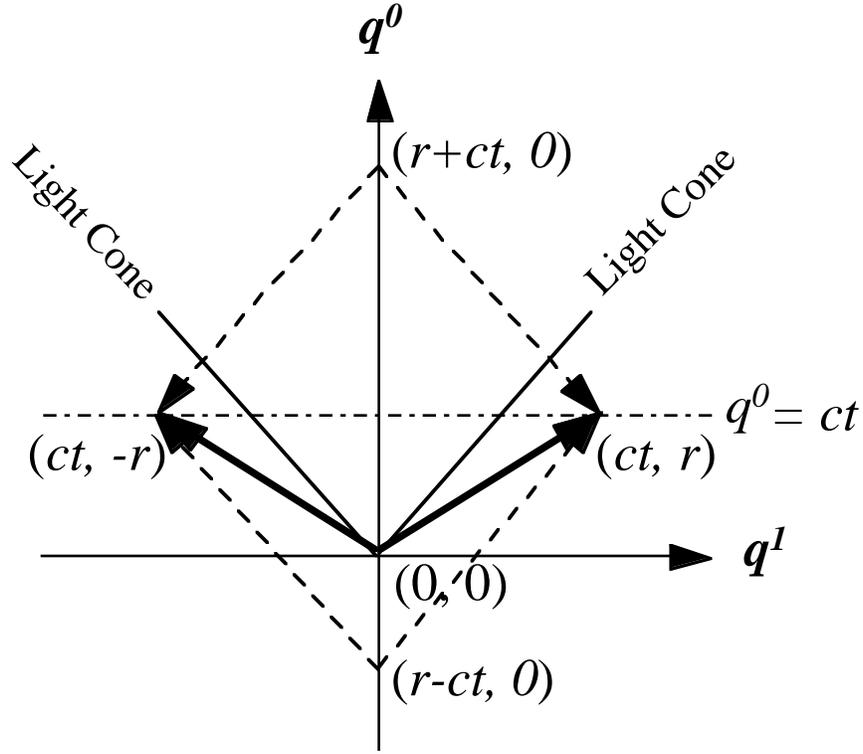

Figure 3. Illustration of the space-time inequality. The Newton-Wigner state $|0, 0_{nw+}\rangle$ represents a particle located at the spatial origin at various times $q^0$ determined by probability distribution $\mathcal{P}(q^0)$. The probability of various position measurement results $\vec{q}$ at time $t$ is determined by probability distribution $\mathcal{P}(t, \vec{q})$. For those times $q^0 < r - ct$ or $q^0 > r + ct$ it is possible for the particle to have traveled a distance $r$ at time $t$ without exceeding the speed of light. Hence, to account for the Hegerfeldt paradox, we need to show that $\mathcal{P}(q^0 = r - ct) + \mathcal{P}(q^0 = r + ct)$ integrated over all $r$ in the interval $(ct, +\infty)$ is greater than $\mathcal{P}(t, |\vec{q}| = r)$ integrated over all $r$ in the same interval.

particle being between $q^0$ and $q^0 + dq^0$. Therefore, the left member of (3.4) is

$$\mathcal{P}(q^0 = r - ct) + \mathcal{P}(q^0 = r + ct) = (2\pi)^{-2}\left[\frac{1}{(r-ct)^2} + \frac{1}{(r+ct)^2}\right]. \qquad (3.7)$$

We wish to compare this with the probability of a position measurement finding the particle at a distance $r$ from the origin at time $t$. This requires that $|0, \vec{0}_{nw+}\rangle$ be expressed in the position representation. We expect the wave function, which we will denote $g(r, t)$, to be spherically symmetric about the spatial origin, because the particle was localized at the origin, has a symmetrical momentum distribution, and has no forces acting on it. The wave function only depends on $r = |\vec{r}|$. Therefore it will be given by an expansion of free spherical waves, satisfying $\vec{r} \times \vec{p} = 0$, normalized so that $\int_0^\infty dr\, 4\pi r^2 |g(r, t)|^2 = 1$, and reducing to $\delta(r)/\sqrt{4\pi}\, r$ as $t \to 0$. We thus obtain[44]

$$g(r, t) = \langle ct, \vec{r}_{loc} | 0, \vec{0}_{nw+}\rangle = \frac{1}{r}\frac{1}{4\pi^{3/2}\hbar}\int_{-\infty}^{+\infty} dp\, \exp\left\{\frac{i}{\hbar}[pr - H(p)t]\right\}, \qquad (3.8)$$

where $H(p) = c\sqrt{p^2 + m^2 c^2}$. [Note: in this subsection, $p$ represents the radial component of $\vec{p}$, not $(p_0, \vec{p})$.] The corresponding *probability density* is

$$\mathcal{P}(t, |\vec{q}| = r) = 4\pi r^2 |g(r, t)|^2. \qquad (3.9)$$

Calculation of the right member of (3.9) is complicated by the square root in the exponential of (3.8). The problem will be approached in two stages. We first take the limit of a massless particle, $m \to 0$, permitting an exact evaluation of $g(r, t)$, and we prove that inequality (3.4) holds in this case. Then, we evaluate the change to $g(r, t)$ as $m$ goes positive and prove the inequality still holds for a massive particle.

### i. The Massless Particle

In the limit $m \to 0$, $H(p) \to c|p|$, and (3.8) becomes

$$g(r, t) = \frac{1}{r}\frac{1}{4\pi^{3/2}\hbar}\int_{-\infty}^{+\infty} dp\, \exp\left[\frac{i}{\hbar}(p\,r - |p|ct)\right]$$

$$= \frac{1}{r} \frac{1}{4\pi^{3/2}\hbar} \left\{ \int_{-\infty}^{0} dp \, \exp\left[\frac{i}{\hbar}p(r+ct)\right] + \int_{0}^{+\infty} dp \, \exp\left[\frac{i}{\hbar}p(r-ct)\right] \right\}$$

$$= \frac{1}{r} \frac{1}{2\pi^{1/2}} \left[ -\frac{i}{2\pi} \mathrm{P}\!\left(\frac{1}{r+ct}\right) + \frac{1}{2}\delta(r+ct) + \frac{i}{2\pi} \mathrm{P}\!\left(\frac{1}{r-ct}\right) + \frac{1}{2}\delta(r-ct) \right]. \quad (3.10)$$

From inspection, as $t \to 0$, $g(r,t) \to \delta(r)/\sqrt{4\pi}\, r$, as expected. Substuting (3.10) into (3.9) we obtain at time $t$ the probability density

$$\mathcal{P}(t, |\vec{q}| = r) = (2\pi)^{-2} \left[ \frac{1}{(r-ct)^2} - \frac{2}{r^2 - c^2 t^2} + \frac{1}{(r+ct)^2} \right]. \quad (3.11)$$

One can see from inspection that (3.11) yields peaks on the light cone $r = \pm ct$. We expect this on physical grounds, because as $m \to 0$, the speed $|v| = |p|/H(p) \to c$, making the light-cone the most likely region to find the particle given $\langle \hat{q}^0 \rangle = 0$.

Now we evaluate the space-time inequality (3.4) for the massless particle. Substituting (3.11) on the right side and (3.7) on the left side of (3.4), we obtain

$$\frac{1}{(r-ct)^2} + \frac{1}{(r+ct)^2} \geq \frac{1}{(r-ct)^2} - \frac{2}{r^2 - c^2 t^2} + \frac{1}{(r+ct)^2} \,. \quad (3.12)$$

Because of the cross terms on the right side, absent on the left, *the inequality is satisfied for m = 0.* Indeed, the cross terms are the result of interference of amplitudes in $g(r,t)$ between contributions from the past and future. This time interference is responsible for Newton-Wigner states being spatially localized in spite of their non-locality in time.

A possible objection to the above analysis is that the time distribution $f(q^0)$ is not normalizable in the sense of ordinary wave functions. However, a normalizable wave function *can* be obtained if we *do not* take the limit $\varepsilon \to 0$ in (3.5). Keeping $\varepsilon$ positive, it can be shown that $\int_{-\infty}^{+\infty} dq^0 |f(q^0)|^2 = (4\pi\varepsilon)^{-1}$, so to normalize $f(q^0)$ we need to multiply it by $\sqrt{4\pi\varepsilon}$. What does this do to the inequality (3.4)? Inspection of (3.10) shows that $g(r,t)$ also is not normalizable, due to singularities at $r = \pm ct$. In fact, for $g(r,t)$ to be normalizable, we have to follow exactly the same procedure, not taking the limit $\varepsilon' \to 0$, and it can be shown that $\int_{-\infty}^{+\infty} dr \left|4\pi r^2 g(r,t)\right|^2 = (4\pi\varepsilon')^{-1}$, yielding a normalization factor for $g(r, t)$ of $\sqrt{4\pi\varepsilon'}$. Since $\varepsilon$ and $\varepsilon'$ are arbitrary, we can choose $\varepsilon = \varepsilon'$ and inequality (3.4) will be satisfied as before.

### ii. The Massive Particle

Now consider the case $m > 0$. This is more difficult, because of the square root appearing in the exponent in (3.8). The theorem on analyticity of Fourier transforms alluded to in Section I (or rather, its converse[9]) will be used to show that the effect of the mass $m$ going positive is to introduce an exponential decay factor in the right member of inequality (3.4), with the result that the inequality is still satisfied.

To simplify calculations, let us define a new radial wave function $h$ in terms of $g$:

$$h(r, t) = \sqrt{4\pi} r \, g(r, t) = (2\pi\hbar)^{-1} \int_{-\infty}^{+\infty} dp \, \exp\left\{\frac{i}{\hbar}\left[pr - H(p)t\right]\right\}, \qquad (3.13)$$

From the form of eq. (3.13), the Fourier transform of $h(r, t)$ is

$$\tilde{h}(p, t) = \exp\left(\frac{i}{\hbar}\sqrt{p^2 + m^2c^2}\, ct\right). \qquad (3.14)$$

The function $\tilde{h}(p, t)$ has an analytic continuation for $p \to p + i\eta$ onto the complex plane in the strip $-m'c \leq \eta \leq m'c$, where $m'$ is a positive number just less than the rest mass, $m$; i.e., $m' + \varepsilon = m$, $\varepsilon$ a positive infinitesimal. A partial binomial expansion yields

$$\frac{H(p + im')}{c} = \sqrt{p^2 + 2ipm'c - m'^2c^2 + m^2c^2} \approx \sqrt{p^2 + 2ipm'c} \approx |p| + ipm'c/|p|. \qquad (3.15)$$

In the approximation $|p| \gg mc$ (valid for most $p$), for positive $p$, $\tilde{h}(p + im', t)$ becomes

$$\tilde{h}(p + im'c, t) = \exp\left(\frac{-i}{\hbar}|p|ct\right)\exp\left(\frac{1}{\hbar}m'c^2t\right), \qquad (3.16)$$

which is a finite quantity (for finite $t$). From the Fourier integral theorem,

$$\tilde{h}(p + im'c, t) = (2\pi\hbar)^{-1/2}\int_{-\infty}^{+\infty} dr \, h(r, t) \exp\left[\frac{-i}{\hbar}(p + im'c)r\right]$$

$$\exp\left(\frac{-i}{\hbar}|p|ct\right)\exp\left(\frac{1}{\hbar}m'c^2t\right) = (2\pi\hbar)^{-1/2}\int_{-\infty}^{+\infty} dr \, h(r, t) \exp\left(\frac{-i}{\hbar}pr\right)\exp\left(\frac{1}{\hbar}m'cr\right)$$

$$\exp\left(\frac{-i}{\hbar}|p|ct\right) = (2\pi\hbar)^{-1/2} \int_{-\infty}^{+\infty} dr\, h(r,t) \exp\left(\frac{-i}{\hbar}pr\right) \exp\left[\frac{m'c}{\hbar}(r-ct)\right]. \quad (3.17)$$

Since the integrand in the right member of the last line in (3.17) contains $h(r,t)$ multiplied by a factor which grows exponentially as a function of $r$ - $ct$, and the expression is finite, $h(r,t)$ *must decay exponentially as* $\exp\left[\frac{-m'c}{\hbar}(r-ct)\right]$.

So we see that the effect of the mass $m$ going positive is to widen the strip of analyticity of $\tilde{h}(p,t)$ on the complex momentum plane, which (taking $m' \to m$) introduces an exponential decay factor of $\exp\left[\frac{-2mc}{\hbar}(r-ct)\right]$ in the probability density $\mathcal{P}(t,|\vec{q}|=r)$ on the right side of inequality (3.4). *This is exactly what we would expect on physical grounds, because the speed* $|v| = c|p|/H(p) = c|p|/\sqrt{p^2+m^2c^2}$ *will be attenuated with increasing mass, yielding a more strongly localized wave function.* The factor $\exp\left[\frac{-2mc}{\hbar}(r-ct)\right]$ is actually a *bound*, since it involves the assumption $|p| \gg mc$. The effect of the approximation is to yield a weaker decay than an exact calculation would yield, because if $|p|$ is smaller, it corresponds to lower velocities, and hence to wave components decreasing faster than this bound.

On the other hand, the exponential factor does not enter into the left side of inequality (3.4), since the rest mass $m$ is in no way involved with the calculation of the time distribution $\mathcal{P}(q^0)$. *Therefore, inequality (3.4) remains satisfied for m > 0.*

This result is entirely consistent with Hegerfeldt's theorem,[4] applied to the case where the wave function is initially exponentially bounded. Hegerfeldt only found "violations of causality" when the probability distribution of the initial state decayed *faster* than $\exp\left[\frac{-2mc}{\hbar}(r-ct)\right]$. Now we see that the decay factor $\exp\left[\frac{-2mc}{\hbar}(r-ct)\right]$ results from the time uncertainty of the state, not superluminal velocities.

### C. Dependence on the Motion of the Observer

We now address the other issue raised in Sec. I, eq. (1.3), that a positive energy state localized in one frame of reference has infinite spatial extent in any reference frame moving with respect to it. It is known that the Newton-Wigner state $\left|x^0, \vec{x}_{nw+}\right\rangle$ is an eigenvector of $\hat{\vec{Q}}(x^0/c)$ with eigenvalue $\vec{x}$. Let us reproduce calculation (3.5) through (3.7) where we performed a position measurement on $\left|0, \vec{0}_{nw+}\right\rangle$, but this time our measurement time is zero; that is, we operate with $\hat{\vec{Q}}(0)$ rather than $\hat{\vec{Q}}(t)$:

$$\hat{\vec{Q}}(0)\left|0, \vec{0}_{nw+}\right\rangle = \int_{-\infty}^{+\infty} dq^0\, \hat{\vec{v}}\left(\frac{-q^0}{c}\right) f(q^0) \left|q^0, \vec{0}_{loc}\right\rangle$$

$$= \frac{-\hat{\vec{v}}}{c} \frac{i}{2\pi} \int_{-\infty}^{+\infty} dq^0 \left|q^0, \vec{0}_{loc}\right\rangle$$

$$= \frac{-\hat{\vec{v}}}{c} \frac{i}{2\pi} \int_{-\infty}^{+\infty} dq^0 \exp\left[\frac{-i}{\hbar} H(\hat{\vec{p}}) q^0\right] \left|0, \vec{0}_{loc}\right\rangle$$

$$= \vec{0}\,. \tag{3.18}$$

Since we have set $t = 0$, when the position eigenvalue $-\vec{v}q^0/c$ is multiplied into the amplitude $i P(1/2\pi q^0) + \delta(q^0)/2$, we obtain $-i\vec{v}/2\pi c$, which is constant with respect to $q^0$. The integral therefore becomes a delta function in $H(\hat{\vec{p}})$ which vanishes because $H(\hat{\vec{p}}) \neq 0$. So in spite of the fact that $\left|0, \vec{0}_{nw+}\right\rangle$ represents the particle being at the origin at different times $q^0$, and has amplitudes for the particle being at distances $\vec{v}q^0/c$ from the origin, a measurement at time $t = 0$ will nevertheless definitely find the particle at $\vec{0}$.

We propose the following explanation of this result. The delta function $\delta(x) = 2\pi^{-1} \int_{-\infty}^{+\infty} dx \exp(-ix)$ is an interference minimum for all values of its argument except zero, where it is a maximum. Its appearance in (3.18) suggests that contributions from various times $q^0$ interfere destructively; thus, localization on the $q^0 = 0$ plane is an interference minimum. Refer again to Fig. 2, recalling that shading denotes regions where the wave function does not vanish. The situation of eq. (3.18) corresponds to a measurement time of $t = 0$; therefore, the solid arrow $(ct, r)$ now lies on the $q^1$ axis. For any given value of $q^0 \neq 0$, another value exists (for a given velocity) such that their contributions precisely cancel. Therefore the only time which contributes to the measurement result is $q^0 = 0$.

But suppose the measurement is taken in a frame of reference moving relativistically in the direction of the $q^1$ axis. The Lorentz transformed spatial axis, denoted in the figure with a tilde ~, is tilted out of the interference minimum. On the $\tilde{q}^1$ axis, cancellation of components from different times does not occur; the transformed wave function $\langle \tilde{q}^0, \tilde{q}^1{}_{loc} | 0, 0_{nw+} \rangle$ is not localized, and cannot vanish for any $\tilde{q}^0$.

### D. Discussion

We will conclude this section with some remarks extending these results to more general quantum states, and a discussion of possible implications for causality.

So far, only point localizations for spinless particles have been considered. But the Hegerfeldt paradox also applies to positive energy states of arbitrary spin having any finite spatial extent, and also to exponentially bounded states. Newton-Wigner states form an orthogonal basis on the physical Hilbert space $\mathcal{H}_{phy}$. Therefore, an arbitrary positive energy state $|\psi_+(x^0)\rangle$ at time $x^0$ can be represented as a superposition

$$|\psi_+(x^0)\rangle = (2\pi m \hbar)^{-1} \int_{-\infty}^{+\infty} d^3x\, \psi_+(x^0, \vec{x}) \int_{-\infty}^{+\infty} d(\lambda \tau) (2\pi \hbar \hat{p}_0)^{1/2} |x_+(\tau)\rangle, \quad (3.19)$$

where expansion (2.36) of the Newton-Wigner state was employed. Every point in the spatial integral of (3.19) for which $\psi_+(x^0, \vec{x})$ is non-vanishing will be surrounded by a position uncertainty, arising from the time uncertainty of its corresponding positive energy augmented state $|x_+(\tau)\rangle = |x^0, \vec{x}_+(\tau)\rangle$ about $x^0$. This leads to the same explanation of the Hegerfeldt paradox for $|\psi_+(x^0)\rangle$ as we found for $|x^0, \vec{x}_{nw+}\rangle$.

If the particle is in an external potential $A^\mu(\hat{q})$, we may obtain exactly the same results if we replace $\hat{p}^\mu$ with $\hat{p}^\mu - \frac{e}{c} A^\mu(\hat{q})$ in all expressions. Finally, we note that all of these results apply equally well to individual spinor components of Dirac particle, since these components are also solutions of the Klein-Gordon equation.[8,34]

Now let us consider whether any new problems with causality have been introduced. The picture that emerges from the forgoing discussion is that Newton-Wigner states $|x^0, \vec{x}_{nw+}\rangle$ appear to be tailor-made to vanish on one space-like hyperplane. As one moves forward along the $q^0$ axis, wave components converge superluminally onto a point, then superluminally diverge.

But nature does not exhibit converging waves; only diverging waves. So if an experimenter can prepare a Newton-Wigner state $|x^0, \vec{x}_{nw+}\rangle$, we must assume that he can do so no sooner that $ct = x^0$. Let us assume for the moment that an ideal measurement prepares such a state. Then from Sec. III-C we know that, at the time, and in the reference frame of its preparation, the only component of the particle's time which contributes to a measurement result is the time of measurement. The particle is "here and now" as far as measurements at that time and in that reference frame are concerned. However the particle still has "virtual" components in the past which can affect measurements in the future or in another reference frame. It may be supposed that those components could concievably influence events leading up to the measurement in such a way as to preclude the measurement's occurrence: this is the "grandfather paradox".

Now let us extend von Neumann's reduction postulate[45] to this model. Since it is the positive-negative energy superposition states $|x^0, \vec{x}_{loc}\rangle$ that appear in the spectral decomposition (2.44), not the Newton-Wigner states $|x^0, \vec{x}_{nw+}\rangle$, we may argue that an infinitely precise measurement with $\hat{\vec{Q}}(x^0/c)$ will yield a state $|x^0, \vec{x}_{loc}\rangle$, not $|x^0, \vec{x}_{nw+}\rangle$. The state $|x^0, \vec{x}_{loc}\rangle$ represents a particle which is definitely at $\vec{q} = \vec{x}$ at the definite time $q^0 = x^0$, with no virtual components in the past. This rules out the "grandfather paradox" as it arises above. It is true that these states have negative energy components moving backwards in time, but this is no more of a problem here than in quantum field theory, where these components are interpreted as anti-particles moving forward in time.

## IV. CONCLUSIONS

In this paper it has been shown that the nonlocality of states demonstrated by the Hegerfeldt paradox has its origin in time uncertainty. It is known that spatially localized states exist for particular times; but without quantizing the time variable, we are left with no quantum mechanical explanation for the infinite expansion of such states within finite time periods, nor for the dependence of their localization on the motion of the observer.

In the parametrization invariant description, on the other hand, a free relativistic particle is constrained in the momenta, but not in position or time, suggesting a temporal freedom which the conventional theory fails to exhibit. Observables appear as constants of the motion, while physical states are superposition of states in an augmented Hilbert space, summed over the proper time. The position observable, acting on a physical state, operates on associated augmented states, allowing measurement results to be interpreted in terms of their properties. Augmented states with strictly positive energy are non-local in time, their temporal distribution being manifested through the position observable as expansion of its spectrum beyond the light cone of what is interpreted as an earlier measurement result. Spatial localization arises on a particular space-like hyperplane through quantum interference over time, position measurements generally receiving contributions from states representing the particle at many times. Since the time of the particle is potentially in the past of the time of measurement, the particle's propagation to distant regions later does not imply superluminal velocities. An inequality can be proven showing that the Hegerfeldt paradox is completely accounted for by the hypotheses of subluminal propagation from a set of initial space-time points determined by the quantum time distribution arising from the positivity of the system's energy. Dependence on the motion of the observer arises because the Lorentz transformation tilts coordinate axes out of the space-like hyperplane wherein the interference minimum is obtained.

## ACKNOWLEDGMENTS

The author wishes to thank C. M. L. Rittby for interesting and fruitful discussions. The author also expresses his gratitude to Juan Leon and Donald H. Kobe for clarifications, and for bringing to the author's attention several important references.

semi-infinite range $[0, +\infty)$, the Feynman propagator would have been obtained instead. This would yield the transition amplitude in only one direction between $x'$ and $x''$, *e.g.*, retarded solutions of the Klein-Gordon equation. A good discussion of these and related Green's functions will be found in Ref. [24], where, in particuar, it is shown that the Hadamard Green's function supports causal propagation, which is consistent with this paper's conclusions.

[43]See, *e.g.*, A. S. Wightman and S. S. Schweber, Phys. Rev. **98**, 812 (1955).

[44]C. Cohen-Tannoudji, B. Diu, and F. Laloë, *Quantum Mechanics*, Vol. II (Hermann and Wiley, Paris, 1977), Ch. 8.

[45]J. von Neumann, *Mathematische Grundlagen der Quantenmechanik* (Springer, Berlin, 1932), pp. 184-237.